\documentclass[12pt]{article}
\usepackage[latin9]{inputenc}
\usepackage[a4paper]{geometry}
\usepackage{float}
\usepackage{mathtools}
\usepackage{amsmath}
\usepackage{amsthm}
\usepackage{amssymb}
\usepackage{cite}
\usepackage{esint}
\usepackage[numbers]{natbib}
\usepackage[unicode=true,
 bookmarks=true,bookmarksnumbered=false,bookmarksopen=false,
 breaklinks=false,pdfborder={0 0 0},pdfborderstyle={},backref=false,colorlinks=false]
 {hyperref}
\hypersetup{pdftitle={Feynman Diagrams},
 pdfauthor={LyX Team, Uwe Stohr, Ronen Abravanel},
 linkcolor=black,citecolor=black,urlcolor=blue,filecolor=blue,pdfpagelayout=OneColumn,pdfnewwindow=true,pdfstartview=XYZ,plainpages=false}

\makeatletter
\numberwithin{equation}{section}
\numberwithin{figure}{section}
\theoremstyle{plain}

\theoremstyle{definition}

\theoremstyle{plain}

\theoremstyle{plain}

\theoremstyle{plain}

\usepackage{braket}
\usepackage{feynmf}
\usepackage{subfigure}

\makeatother

\providecommand{\corollaryname}{Corollary}
\providecommand{\definitionname}{Definition}
\providecommand{\lemmaname}{Lemma}
\providecommand{\propositionname}{Proposition}
\providecommand{\theoremname}{Theorem}

\begin{document}


\begin{titlepage}

\thispagestyle{empty}

\begin{flushright}
\end{flushright}

\vspace{.4cm}
\begin{center}
\noindent{\large \bf Generalizations of Reflected Entropy and\\ the Holographic Dual}\\
\vspace{2cm}

Jinwei Chu, Runze Qi, Yang Zhou
\vspace{1cm}

{\it
Department of Physics and Center for Field Theory and Particle Physics, \\
Fudan University, Shanghai 200433, China\\
}

\end{center}

\vspace{.5cm}
\begin{abstract}
We introduce a new class of quantum and classical correlation measures by generalizing the reflected entropy to multipartite states. We define the new measures for quantum systems in one spatial dimension. For quantum systems having gravity duals, we show that the holographic duals of these new measures are various types of minimal surfaces consist of different entanglement wedge cross sections. One special generalized reflected entropy is $\Delta_R$, with the holographic dual proportional to the so called multipartite entanglement wedge cross section $\Delta_W$ defined before. We then perform a large $c$ computation of $\Delta_R$ and find precise agreement with the holographic computation of 2$\Delta_{W}$. This agreement shows another candidate $\Delta_R$ as the dual of $\Delta_W$ and also supports our holographic conjecture of the new class of generalized reflected entropies.
\end{abstract}

\end{titlepage}

\newpage
\section{Introduction}

The goal of this paper is to propose a large class of new correlation measures for multipartite systems and their holographic duals, motivated by the recent work~\cite{Dutta:2019gen}.
In most of the current studies of quantum entanglement~\cite{Vidal:2002rm,Kitaev:2005dm,Levin:2006,Casini:2004bw,Casini:2012ei,Casini:2008cr,Hung:2018rhg,RT:RT-formula,HRT:HRT-formula} we often
divide the total system into two: $A$ and its complement $A^{c}$, and compute the entanglement entropy $S_{A}:=-{\rm Tr}\rho_{A}\log\rho_{A}$.
In the gauge/gravity correspondence
\citep{Maldacena:AdS/CFT}, the Ryu-Takayanagi formula \citep{RT:RT-formula,HRT:HRT-formula}
allows us to use a codimension-2 surface in AdS to compute the entanglement entropy in holographic CFTs. This discovery starts a new era
of studying relations between spacetime geometry and quantum
entanglement precisely~\citep{Swingle:TensorNetwork,VanRaamsdonk:Buildingup1,MaldacenaSusskind:ERequalsEPR,FGHMR:LinearEinsteinequationfrom1stLaw,MiyajiTakayanagi:SurfaceStateCorrespondence,MTW:OptimizationofPathIntegral1,CKMTW:OptimizationofPathIntegral2,PYHP:HaPPYcode,FreedmanHeadrick:BitThreads}.

It has been known that the entanglement entropy truly measures quantum entanglement
only for pure states $\ket{\psi}_{AB}$. Therefore it is interesting to ask what is the analogy of that and the Ryu-Takayanagi formula when $\rho_{AB}$ is a mixed state. Recently there have been several proposals~\cite{Dutta:2019gen, Kudler-Flam:2018qjo,Tamaoka:2018ned,Kudler-Flam:2019oru,Kudler-Flam:2019wtv,Kusuki:2019zsp} and it turns out that the entanglement wedge cross section wins most of the attentions in the dual gravity side.
Another important question is how to find multipartite correlation measures and their geometric dual. It is known that there are much richer correlation structures in quantum
systems consisting of three or more subsystems (see e.g. \citep{HHHH:QuantumEntanglement}). However the holographic interpretation of multipartite correlations is less known, though it is obviously crucial for the understanding of the emergence of bulk geometry from many-body quantum entanglement on the boundary.

In~\cite{Umemoto:2018jpc}, an analogy of bipartite entanglement wedge cross section for multiple subsystems, $\Delta_W$ has been proposed and it shares a lot of common features with the multipartite generalization of entanglement of purification $\Delta_P$. This motivates the authors in~\cite{Umemoto:2018jpc} to propose the conjecture $\Delta_P=\Delta_W$. The difficulty to compute $\Delta_P$ makes it a bit hard to test $\Delta_P=\Delta_W$ though this conjecture is the most natural generalization of the $E_P=E_W$ conjecture proposed in~\cite{Takayanagi:2017knl,NDHZS:HEoP}. See~\cite{Bao:2018gck,Takayanagi:2018zqx,Cui:2018dyq,Yang:2018gfq,Bao:2018fso,Agon:2018lwq,Heydeman:2018qty,Caputa:2018xuf,Guo:2019azy,Liu:2019qje,Bhattacharyya:2019tsi,Ghodrati:2019hnn,BabaeiVelni:2019pkw,Bao:2019bib,Du:2019emy,Guo:2019pfl,Bao:2019wcf,Harper:2019lff,Akhavan:2019zax,Kusuki:2019rbk,Zhou:2019jlh,Umemoto:2019jlz,Jeong:2019xdr,Kusuki:2019evw,CKNR:EntanglementWedge,Wall:Maximinsurfaces,HHLR:EntanglementWedge,Bao:2019zqc,Kusuki:2019evw} for recent progress.

In this paper, we introduce a new class of multipartite correlation measures in generic quantum systems.
They are defined by generalizing the reflected entropy method for multipartite systems. Among these new measures, there is a special one called multipartite reflected entropy $\Delta_R$ invariant under the permutations of subsystems.

We then show that the holographic duals of our new measures are different types of minimal codimension-2 surfaces in the entanglement wedge \citep{CKNR:EntanglementWedge,Wall:Maximinsurfaces,HHLR:EntanglementWedge}, motivated by Ryu-Takayanagi proposal for multi-boundaries. In particular the holographic dual of $\Delta_R$ is proportional to $\Delta_W$ defined in~\cite{Umemoto:2018jpc}. We perform the large $c$ computation of $\Delta_R$ using replica trick and twist operators and find precise agreement with the holographic computation in AdS$_3$/CFT$_2$. This agreement tempts us to propose another candidate dual to multipartite entanglement wedge cross section $\Delta_{R}=2\Delta_{W}$ and also strongly supports our holographic conjectures for the new class of generalized reflected entropies.

This paper is organized as follows: 
In Section \ref{sec2}, we give the definition of a class of generalized reflected entropies and focus on a special one $\Delta_R$ invariant under permutations of subsystems.
In Section \ref{sec3}, we introduce a class of multipartite generalizations of entanglement wedge cross-section in holography and find that there is a one to one correspondence with the generalizations of reflected entropy.
In Section \ref{sec4}, we perform a large $c$ computation of $\Delta_R$ in tripartite case and find agreement with holographic computation. This agreement supports our holographic conjectures between generalized reflected entropies and generalized entanglement wedge cross-sections. We discuss some information theoretic properties of $\Delta_R$ in Section \ref{sec5} and conclude in Section \ref{sec6}.

\textit{Note added}: After all the results in this paper were obtained,  \citep{Bao:2019zqc} appeared in which they construct similar
generalization of reflected entropy for $\Delta_W$, which is different from ours.

\section{Generalized reflected entropy\label{sec2}}

\begin{figure}[htbp]
	\centering
	\includegraphics[width=4in]{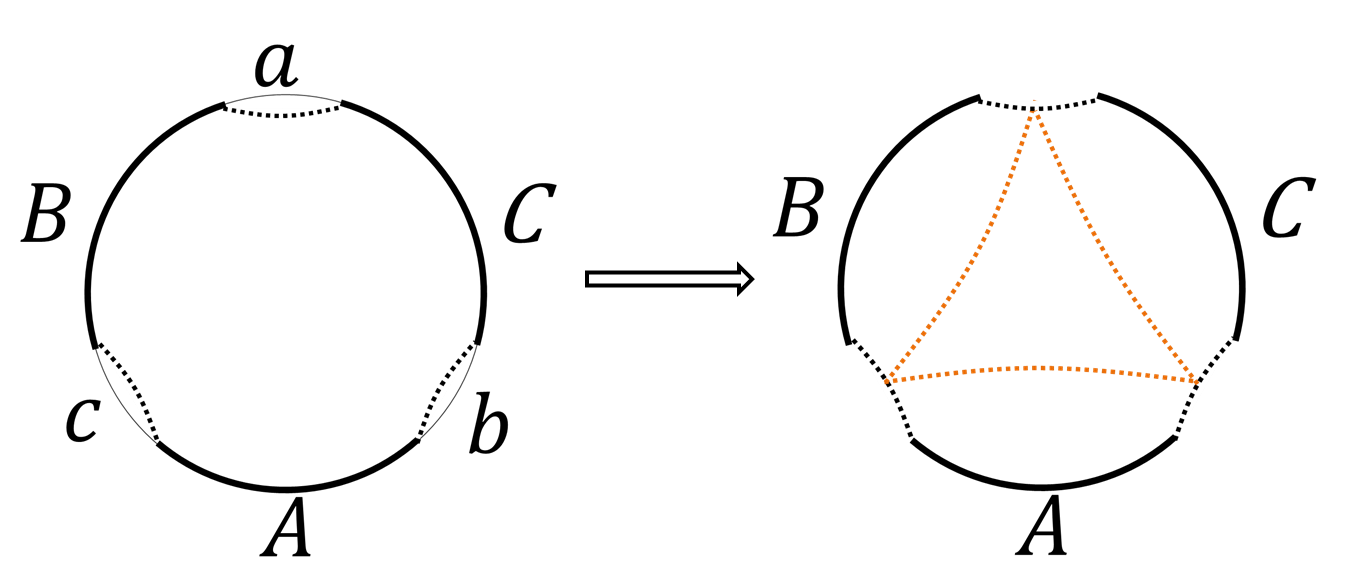}\\
	\caption{Tripartite entanglement wedge cross-sections $\Delta_W$ of subsystems $ABC$ in AdS$_3$/CFT$_2$. $\mathit{Left}:$ A pure state in $2d$ CFT on a circle made up of six intervals: $A,B,C,a,b\text{ and }c$. The dotted lines denote Ryu-Takayanagi surfaces of $ABC$. $\mathit{Right}:$ Entanglement wedge, the interior of the Ryu-Takayanagi surfaces $\cup\  ABC$, in which the closed curve denotes $\Delta_W$.}
	\label{ew}
	\end{figure}

Consider a quantum state on a circle, which is made up of six intervals: $A,B,C,a,b$ and $c$, shown in Fig.\ref{ew}. For holographic CFTs, it is known that the holographic entanglement entropy for $\rho_{ABC}$ is given by the Ryu-Takayanagi surface, the sum of 3 bulk geodesics bounded by the ends of $A,B,C$, as shown in Fig.\ref{ew}. Recently the triangle type of 3 other geodesics, with 3 ends located on the bulk Ryu-Takayanagi surfaces, has been defined as the multipartite entanglement wedge cross-sections, $\Delta_W$~\cite{Umemoto:2018jpc}. This has been understood as a total correlation measure among subsystems $A$, $B$ and $C$. In particular, the triangle sum can not be decomposed as (sum of) bipartite entanglement wedge cross-sections. This suggests that the measure $\Delta_W$ is an intrinsic 3-body correlation measure. One obvious question is how to understand the triangle from CFT point of view. This is one of our motivations.
	
In this section, we define a class of correlation measures for $1+1$ dimensional quantum field theory (QFT) state on a circle. The following definition was motivated by a holographic CFT state on a circle in AdS$_3$/CFT$_2$. But we stress that the definition itself is independent of holography. 
Recently an interesting measure called {\it reflected entropy} has been proposed as a bipartite correlation measure for a mixed state $\rho_{AB}$~\cite{Dutta:2019gen}. The idea is to introduce a canonical purification for subsystems $A,B$ and then measure the entanglement entropy. 
Without going into the detail definition of reflected entropy in information theory, let us understand reflected entropy in the following intuitive way. Start from a pure state $\psi_{ABc}\in\mathcal{H}_{ABc}$ defined on a circle and the mixed state $\rho_{AB}$ can be viewed as the reduced density matrix by tracing out $c$. There is a simple and canonical purification for a given $\rho_{AB}$ by doubling the Hilbert space:
\begin{equation}
|\sqrt{\rho_{AB}}\rangle = |\sqrt{\text{Tr}_c|\psi\rangle\langle\psi|}\rangle \in (\mathcal{H}_{A}\otimes\mathcal{H}_{A^*})\otimes(\mathcal{H}_{B}\otimes\mathcal{H}_{B^*})\equiv \mathcal{H}_{AA^*BB^*}\ .
\end{equation} This can be obtained by flipping Bras to Kets for basis of a given density matrix $\rho_{AB}$. It can be shown that 
\begin{equation}
\rho_{AB} = \text{Tr}_{\mathcal{H}_{A^*B^*}} |\sqrt{\rho_{AB}}\rangle\langle\sqrt{\rho_{AB}}|\ .
\end{equation} The reflected entropy is defined as 
\begin{equation}
S_R(A:B) := S(AA^*)_{\sqrt{\rho_{AB}}}\ .
\end{equation}
The reflected entropy turns out to be a good measure of correlations between $A$ and $B$ for state $\rho_{AB}$~\cite{Dutta:2019gen}:
\begin{equation}
\text{pure state}:~S_R(A:B) = 2S(A)\ ,
\end{equation}
\begin{equation}
\text{factorized state}:~S_R(A:B) = 0\ ,
\end{equation}
\begin{equation}
\text{bounded from below}:~S_R(A:B) \geq I(A:B)\ ,
\end{equation}
\begin{equation}
\text{bounded from above}:~S_R(A:B) \leq 2\text{min}\{S(A),S(B)\}\ ,
\end{equation}
\begin{equation}
\text{for states saturating Araki-Lieb inequality}:~S_R(A:B)= 2\text{min}\{S(A),S(B)\}\ .
\end{equation}
\begin{figure}[t]
\centering{}\includegraphics[scale=0.3]{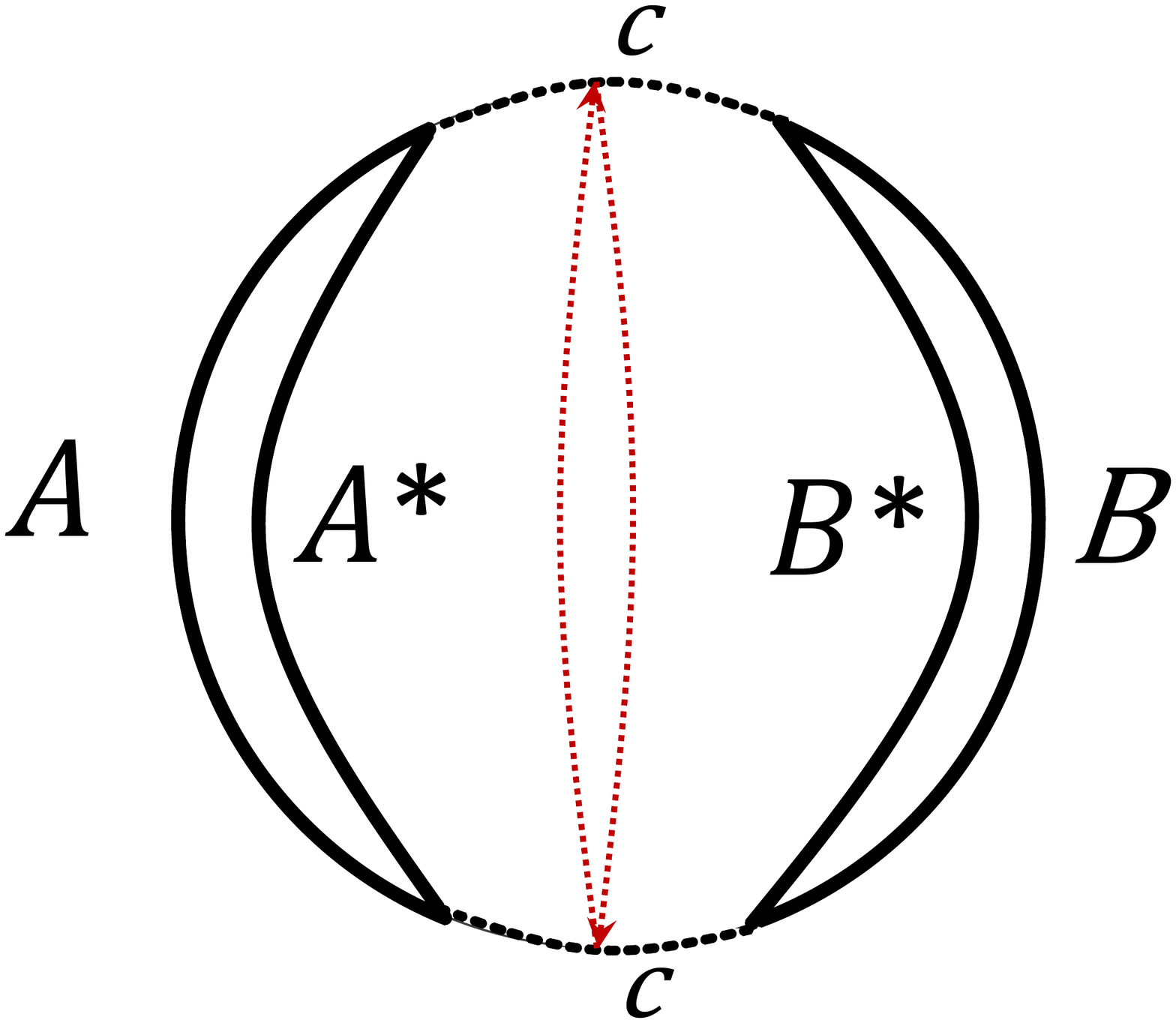}\caption{\label{fig:MER1} Canonical purification of $\rho_{AB}$: $|\sqrt{\rho_{AB}}\rangle = |\sqrt{\text{Tr}_c|\psi\rangle\langle\psi|}\rangle$. Tracing out $c$ corresponds to gluing $c$ from 2 circles and we view this process as a fundamental step to obtain a big pure state. The red dashed line separates $AA^*$ from $BB^*$ and defines reflected entropy $S_R$.
}
\end{figure}

Let us give a graph description of the canonical purification procedure in Fig.\ref{fig:MER1}. Assign a circle for each Hilbert space. Start from the pure state $\psi_{ABc}$ and glue $c$ from the two circles and we obtain the purified state
\begin{equation}
|\sqrt{\rho_{AB}}\rangle = |\sqrt{\text{Tr}_c|\psi\rangle\langle\psi|}\rangle\ .
\end{equation}
This should be viewed as a fundamental step to build up another canonical pure state start from one pure state. We stress that the final canonical state is independent of $c$, because for a given $\rho_{AB}$ one can choose another $c'$ which does the same purification as $c$ and the final canonical state would not change. Therefore the reflected entropy is independent of $c$. This is not surprising because $S_R$ is an intrinsic property of the mixed state $\rho_{AB}$. Later we will see that $c$ is helpful to understand the global structure when we have a big complicated purified state. This is roughly because a nontrivial $c$ in our setup indicates that the initial state $\rho_{AB}$ is a mixed state or in another word $AB$ is entangled with others and we do not know the full information of $AB$. Related to this, after gluing along $c$, one can schematically view $c$ representing some entanglement between $AB$ and $A^*B^*$. Another convenient way to understand Fig.\ref{fig:MER1} is to imagine that there are $2d$ spacetime surfaces bounded by circles. The possible meaning of the radial direction is Euclidean time. Consider all the states in the formalism of path integral. After gluing two spacetime patches along $c$ we have obtained a pure state associated to two boundaries $AA^*$ and $BB^*$. The red curve along two spacetime patches readily separates two boundaries $AA^*$ and $BB^*$ and plays the role of the entangling surface in spacetime. After all these constructions and interpretations we can define a robust entanglement entropy associated to the red curve, the reflected entropy
\begin{equation}
S_R(A:B)=S(AA^*:BB^*)_{\sqrt{\rho_{AB}}}=\text{Entanglement Entropy of Red Curve}\ .
\end{equation}

Now we are ready to generalize our construction of canonical purification to multipartite $\rho_{ABC\cdots}$. Consider a state on a circle made up of six intervals: $A,B,C,a,b\text{ and }c$, shown in Fig.\ref{ew}. We can do different canonical purifications by gluing different regions $a$, $b$ or $c$. 

The easiest way is to pick up two circles and glue $a,b,c$ once and we get a pure state $\sqrt{\rho_{ABC}}$. Since the spacetime geometry after gluing is like a pair of pants, one can have 3 options to draw a red curve to separate 3 boundaries $AA^*$, $BB^*$ and $CC^*$ respectively from other parts. These correspond to measure the reflected entropy for bipartitions $(A:BC)$, $(B:AC)$ and $(C:AB)$
\begin{equation}
S_R(A:BC)=S(AA^*:BB^*CC^*)\sqrt{\rho_{ABC}}\ ,
\end{equation} 
\begin{equation}
S_R(B:AC)=S(BB^*:AA^*CC^*)\sqrt{\rho_{ABC}}\ ,
\end{equation} 
\begin{equation}
S_R(C:AB)=S(CC^*:AA^*BB^*)\sqrt{\rho_{ABC}}\ .
\end{equation} 

One can also perform 2 steps of canonical purification to create a pure state using 4 copies of $\mathcal{H}_{ABC}$. For instance, we first perform the canonical purification by gluing $c$ from two copies $\mathcal{H}_{ABCabc}$ and $\mathcal{H}_{A'B'C'a'b'c'}$ and obtain
\begin{equation}
\label{stp1}
\psi_1=|\sqrt{\text{Tr}_c|\psi_{ABCabc}\rangle\langle\psi_{ABCabc}|}\,\rangle\ .
\end{equation}
Then we pick up another copy of $\psi_1$ and do canonical purification again by gluing $b$ and $b'$  and obtain
\begin{equation}
\label{stp2}
\psi_2=|\sqrt{\text{Tr}_{bb'}|\psi_1\rangle\langle\psi_1|}\,\rangle\ .
\end{equation}
Now we are left with $a,a',a'',a'''$ and we can pair them and glue. We can try to draw red curves to bipartition the final pure state in the Hilbert space consisting of 4-copy of $\mathcal{H}_{ABC}$. Entanglement entropy of each curve will measure some correlations among $\rho_{ABC}$. These will include some biparitite reflected entropy detected in the 2-copy purification mentioned before and also some other new measures.

	\begin{figure}[htbp]
	\centering
	\includegraphics[width=6in]{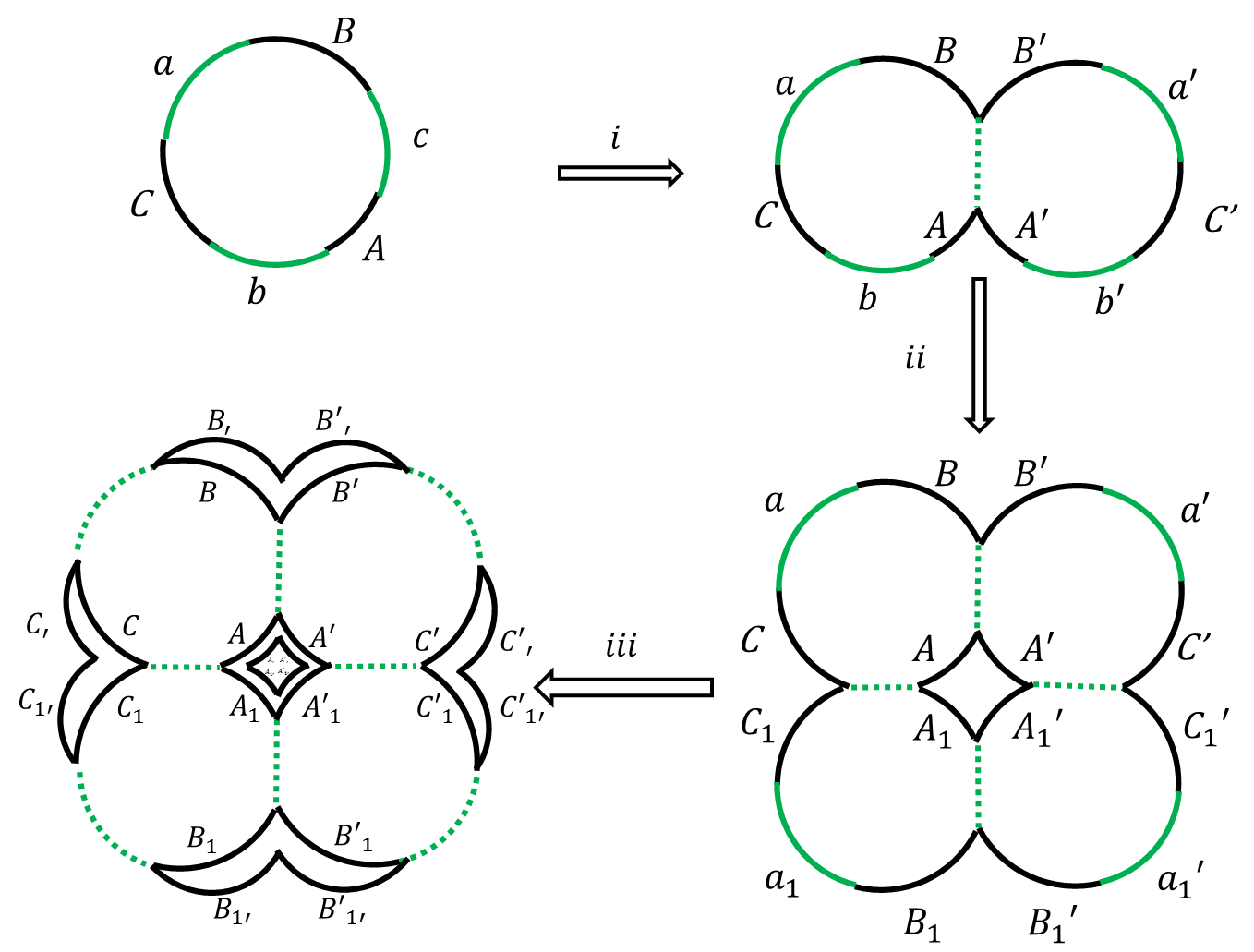}\\
	\caption{The procedure to construct the pure state with three similar steps. Step $i$: from the original pure state $\rho_0=|\psi_{ABCabc}\rangle \langle \psi_{ABCabc}|$ to $\psi_1=|\sqrt{\text{Tr}_c\rho_0}\,\rangle$. Step $ii$: from $\rho_1=|\psi_1\rangle \langle \psi_1|$ to $\psi_2=|\sqrt{\text{Tr}_{bb'}\rho_1}\,\rangle$. Step $iii$: from $\rho_2=|\psi_2\rangle \langle \psi_2|$ to $\psi_3=|\sqrt{\text{Tr}_{aa'a''a'''}\rho_2}\,\rangle$, whose density matrix is $\rho_3=|\psi_3\rangle \langle \psi_3|$ and this is the boundary state in final 8-copy purification (also seen in Fig.\ref{ps}).}
	\label{ptc}
	\end{figure}	

In this work we are particularly interested in another purification involving $8$ copies of $\mathcal{H}_{ABC}$ for the reason we will see later. By adding one more step of canonical purification to the 4-copies purification by doubling Hilbert space one can get 
\begin{equation}
\label{stp3}
\psi_3=|\sqrt{\text{Tr}_{aa'a''a'''}|\psi_2\rangle\langle\psi_2|}\,\rangle\ .
\end{equation}
In order to make it more transparent we draw our purification process in Fig.\ref{ptc}. We switch our notations a little bit for labeling different copies. We stress that even though $a,b,c$ (and the copies of them) are involved in the purification process, the final big pure state $\psi_3$ does not depend on $a,b,c$ and their copies because essentially all of them are traced out. To understand this better, one can view $a,b,c$ as a certain purification for $\rho_{ABC}$ in the beginning and change them to another purification will not affect the final big state constructed here. According to the notation in Fig.\ref{ptc} the final state involves 8 copies of $A,B,C$ and it should be denoted specifically as 
\begin{equation}
\psi_3=\psi_{AA'A_1A'_1A_{'}A'_{'}A_{1'}A'_{1'}BB'B_{'}B'_{'}B_1B'_1B_{1'}B'_{1'}CC_{'}C_1C_{1'}C'C'_{'}C'_1C'_{1'}}\ .
\end{equation} 
One can now try to draw curves to bipartition the final pure state $\psi_3$. There are certain curves running over all bridges among $a,b,c$. For instance, one such curve separates the big pure state into two and the entanglement entropy associated with that curve is given by
\begin{equation}
\label{bp}
\begin{split}
\Delta_{R}(A:B:C)\equiv S(AA'A_1A'_1B_1B'_1B_{1'}B'_{1'}CC_{'}C_1C_{1'}:A_{'}A'_{'}A_{1'}A'_{1'}BB'B_{'}B'_{'}C'C'_{'}C'_1C'_{1'})_{\psi_3}\ .
\end{split}
\end{equation}
We define such entanglement entropy as {\it multipartite reflected entropy}. We stress again that for each curve doing the bipartition there is a well defined generalized reflected entropy.

Last but not least, for any given pure state constructed by the above procedure, one can trace out some part of it and get a new mixed state. And one can do once more canonical purification for this mixed density matrix and obtain another new pure state.
It is not hard to realize that by such kinds of constructions, we can build a pure state in any even number copies of Hilbert spaces. 

We can compute these entropies using replica trick. For instance, as the $\pmb{n}\to 1$ limit of R\'enyi entropy $\Delta_R$ can be computed by 
\begin{equation}
\label{ry1}
\begin{split}
\Delta_R(A:B:C)=\lim_{\pmb{n} \to 1}S_{\pmb{n}}, \quad S_{\pmb{n}}=\frac{1}{1-\pmb{n}}\ln\text{Tr}_R(\text{Tr}_L\rho_3)^{\pmb{n}}
\end{split}
\end{equation} 
where $L$ denotes the left side of the bi-partition in (\ref{bp}), namely
\begin{equation}
\begin{split}
L\equiv \{ AA'A_1A'_1, B_1B'_1B_{1'}B'_{1'},CC_{'}C_1C_{1'}\}\ .
\end{split}
\end{equation}

\section{Holography of generalized reflected entropy\label{sec3}}
In the previous section, we construct many big pure states by performing canonical purifications for a quantum system on a circle and define different generalized reflected entropies from them. Though some of the newly defined entropies are inspired from holography, all the definitions by themselves are independent of holography. We study their holographic duals in this section. Again we will first understand the bipartite case and the multipartite generalization will be understood straightforwardly after that. The bipartite case has been largely developed in~\cite{Dutta:2019gen}. We review the bipartite case for the purpose of generalizations. Notice that previously we perform canonical purification by solely working with quantum systems on a circle. Now for those quantum systems having bulk gravity dual, we have to extend the previous gluing procedure together with the bulk. For simplicity we will focus on static cases through this section.

\begin{figure}[t]
\centering{}\includegraphics[scale=0.3]{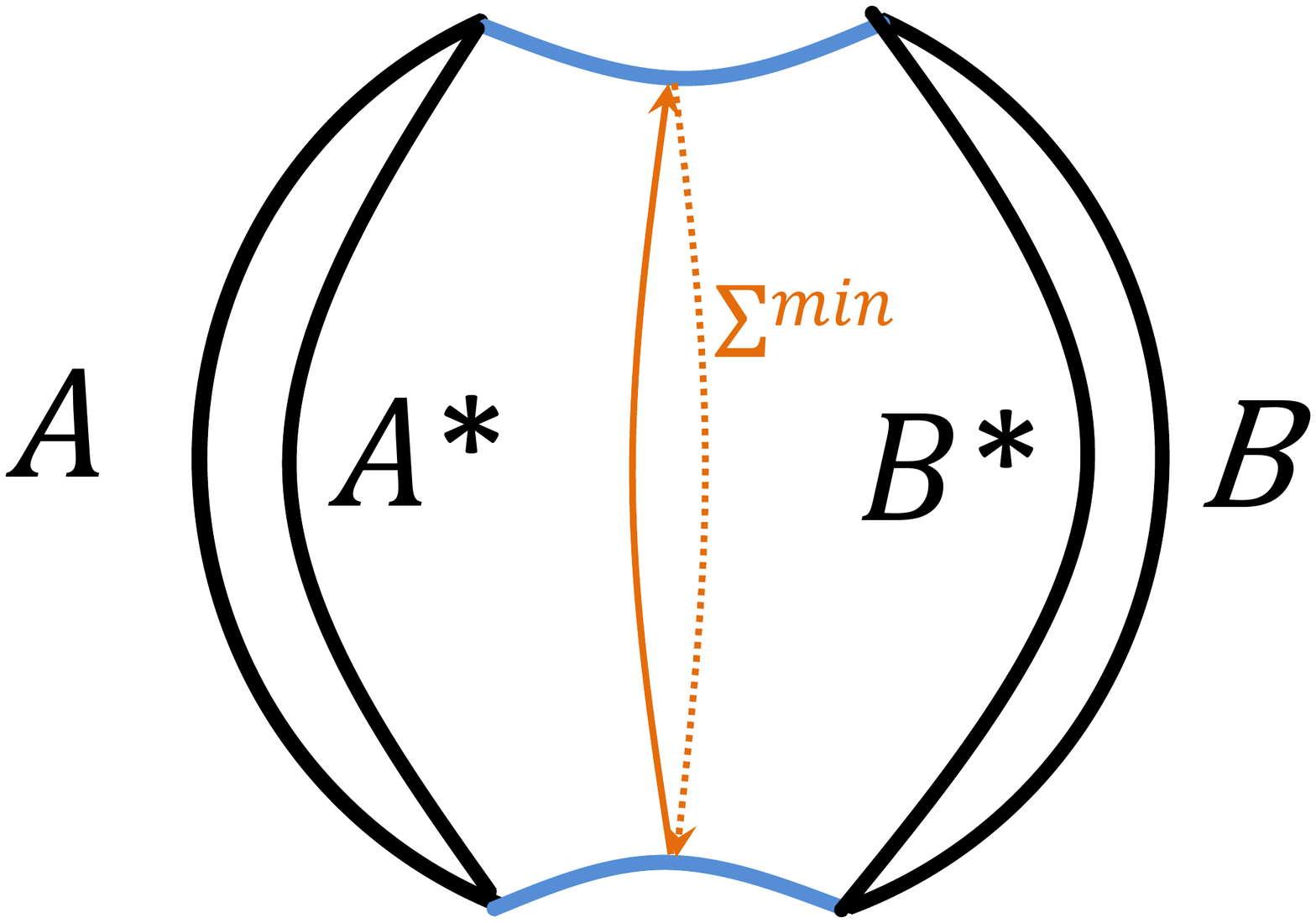}\caption{\label{fig:MER2} Canonical purification of $\rho_{AB}$ together with entanglement wedges: Tracing out $c$ corresponds to gluing Ryu-Takayanagi surfaces (blue lines) for two copies of entanglement wedges and the new bulk geometry describes two entangled boundary quantum systems $AA^*$ and $BB^*$. We view this process as a fundamental step to obtain a bulk geometry describing a big pure state. The orange line is the minimal surface in the bulk seperating $AA^*$ from $BB^*$.
}
\end{figure}

Let us first recall the case of $\rho_{AB}$. Start from a global pure state $\psi_{ABc}$ having a classical bulk solution as its gravity dual. Tracing out $c$ corresponds to discard other bulk regions and keep only the entanglement wedge for $\rho_{AB}$. For a fixed time slice, this was defined as the region bounded by $A\cup B\cup \Gamma_{AB}$ where $\Gamma_{AB}$ is the Ryu-Takayanagi surfaces for $\rho_{AB}$. Now doubling the Hilbert space for $\mathcal{H}_{AB}$ means to pick up another copy of the entanglement wedge. Doing the canonical purification for boundary $\rho_{AB}$ would correspond to gluing the bulk entanglement wedges along $\Gamma_{AB}$ since this is the most natural way to construct the new bulk geometry to respect the purified boundary constructed in the previous section without creating new boundaries. We draw the constructed bulk geometry in Fig.~\ref{fig:MER2}.

Now the question is which minimal surface is the geometrical dual of the reflected entropy $S_R(A:B)$ constructed in the previous section, the entanglement entropy between $AA^*$ and $BB^*$ for the constructed pure state. Intuitively this surface (line in the present example) should count the entanglement flux between $AA^*$ and $BB^*$ and is naturally given by the so called entanglement wedge cross section on each copy. Because of a $\mathbb{Z}_2$ symmetry under exchanges of $A$ and $A^*$, $B$ and $B^*$, the geometry dual of $S_R(A:B)$, the closed minimal curve $\Sigma^{min}$ is exactly twice of $E_W$. This is one of the main results in~\cite{Dutta:2019gen}, where a number of evidences have been provided to support this duality.

Let us now do some comparison with Fig.\ref{fig:MER1} since they are closely related. In Fig.\ref{fig:MER1}, dashed black lines $c$ do not correspond to real physical objects. They just indicate which part we have traced out. And the dashed red curve there does not correspond to any physical object either. They are used to bipartition the quantum system described by the final pure state. Here things are rather different. Both the real blue lines and the orange lines have precise physical meanings. The former is the Ryu-Takayanagi surfaces for the entanglement entropy of $\rho_{AB}$ and the latter is the geometric dual of the reflected entropy. 
Before we generalize the above geometric dual of reflected entropy to multi-partite cases, let us give some general remarks. In previous section, we associated a general reflected entropy to each curve separating the final pure state into two. For theories having classical bulk duals, due to similar geometric structures there will be one to one correspondence between the separating curve in the previous section and the minimal curve in this section. Therefore we expect for any well constructed generalized reflected entropy there will be a minimal surface dual to it. Let us stress that this argument will lead us to find duality between a large class of generalized reflected entropies and new types of minimal surfaces consist of entanglement wedge cross sections, beyond those known before~\cite{Umemoto:2018jpc,Takayanagi:2017knl,NDHZS:HEoP}.

\begin{figure}[htbp]
\centering
\includegraphics[width=5in]{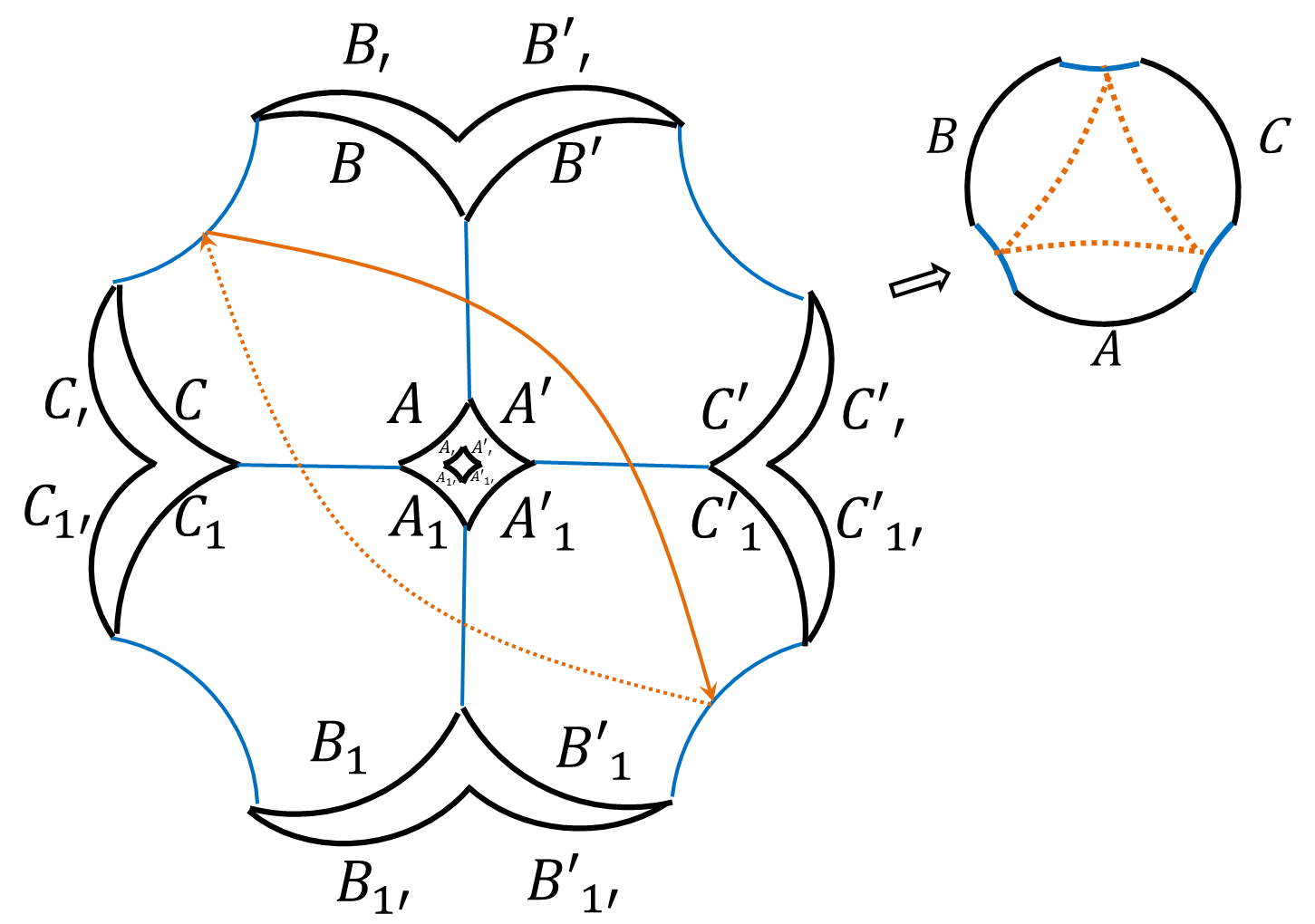}\\
\caption{A pure state constructed by 8 copies of the subsystem $ABC$ together with dual glued bulk. The entangling surface (denoted by the closed orange curve) is just twice of the minimal cross sections in Fig.\ref{ew}, which is also the holographic dual of entanglement entropy $S(AA'A_1A'_1B_1B'_1B_{1'}B'_{1'}CC_{'}C_1C_{1'}:A_{'}A'_{'}A_{1'}A'_{1'}BB'B_{'}B'_{'}C'C'_{'}C'_1C'_{1'})$, defined to be multipartite reflected entropy of subsystems $ABC$, namely $\Delta_R(A:B:C)$. It can be seen that $\Delta_R(A:B:C)=2\Delta_W(A:B:C)$ for holographic states.}
\label{ps}
\end{figure}	

Now we are ready to generalize the canonical purification procedure together with the bulk to multipartite cases. Let us first discuss 3-body mixed state $\rho_{ABC}$ defined on a circle. One can essentially repeat what we discussed in last section for 2-copy purification, 4-copy purification, 8-copy purification by adding the bulk. Without further analysis, let us list the dual reflected entropy for different types of minimal surfaces constructed in the bulk below. We particularly draw the final big pure state in 8-copy canonical purification in Fig.\ref{ps} where the orange line denotes a minimal curve which is the bulk geometric dual to multipartite reflected entropy $\Delta_R(A:B:C)$ constructed in previous section. It is easy to see that this is twice of the multipartite entanglement wedge cross sections $\Delta_W(A:B:C)$ defined in~\citep{Umemoto:2018jpc}.
\begin{figure}[htbp]
\centering
\includegraphics[width=4in]{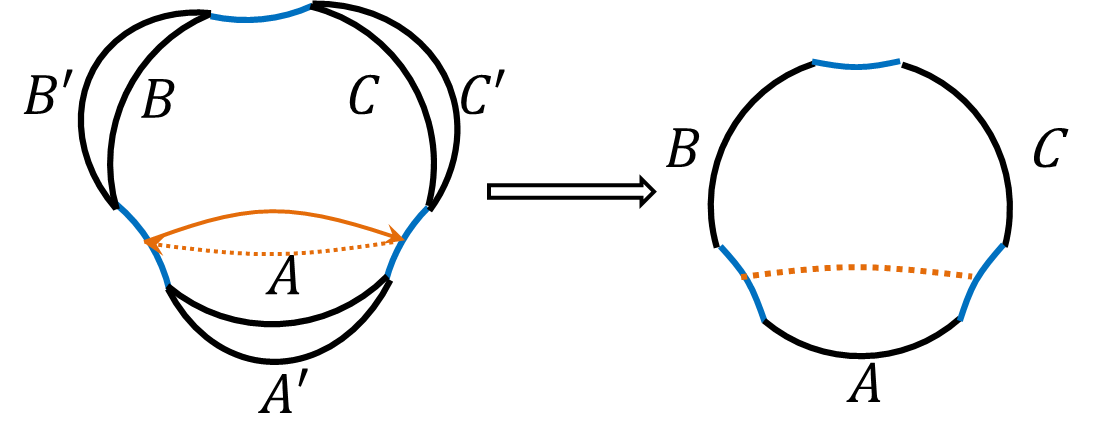}\\
\caption{Canonical purification of $\rho_{ABC}$ ($\mathit{left}$) with the minimal cross section denoted by the orange line dual to $S_R(A:BC)$. This is twice of $E_W(A:BC)$ denoted by the orange line in the entanglement wedge of $\rho_{ABC}$ ($\mathit{right}$)}
\label{2copy}
\end{figure}

Similarly, we draw a pure state in 2-copy canonical purification in Fig.\ref{2copy} where the left orange line denotes a minimal curve which is dual to reflected entropy $S_R(A:BC)$. It can be seen that this is twice of bipartite cross-section $E_W(A:BC)$.
\begin{figure}[htbp]
\centering
\includegraphics[width=5in]{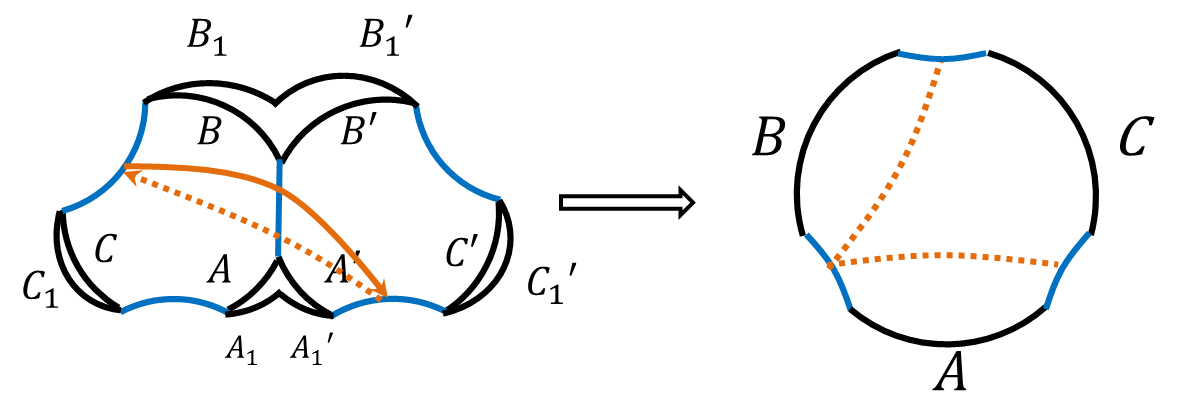}\\
\caption{Canonical purification of $\rho_1$ ($\mathit{left}$) with the minimal cross section denoted by the orange line dual to $S(AA'A_1A_1'CC_1)_{|\sqrt{\rho_1}\rangle}$. This is twice of $\Sigma_{(1)}^{min}(C:A:B)$ denoted by the orange line in the entanglement wedge of $\rho_{ABC}$ ($\mathit{right}$).}
\label{4copy}
\end{figure}

Then, we draw a pure state in 4-copy canonical purification in Fig.\ref{4copy} where the left orange line denotes a minimal curve which is dual to $S(AA'A_1A_1'CC_1)_{|\sqrt{\rho_1}\rangle}$. It can be seen that this is twice of $\Sigma_{(1)}^{min}(C:A:B)$ defined as the minimal curve with the shape shown in right figure of Fig.\ref{4copy}.
\begin{figure}[htbp]
\centering
\includegraphics[width=6in]{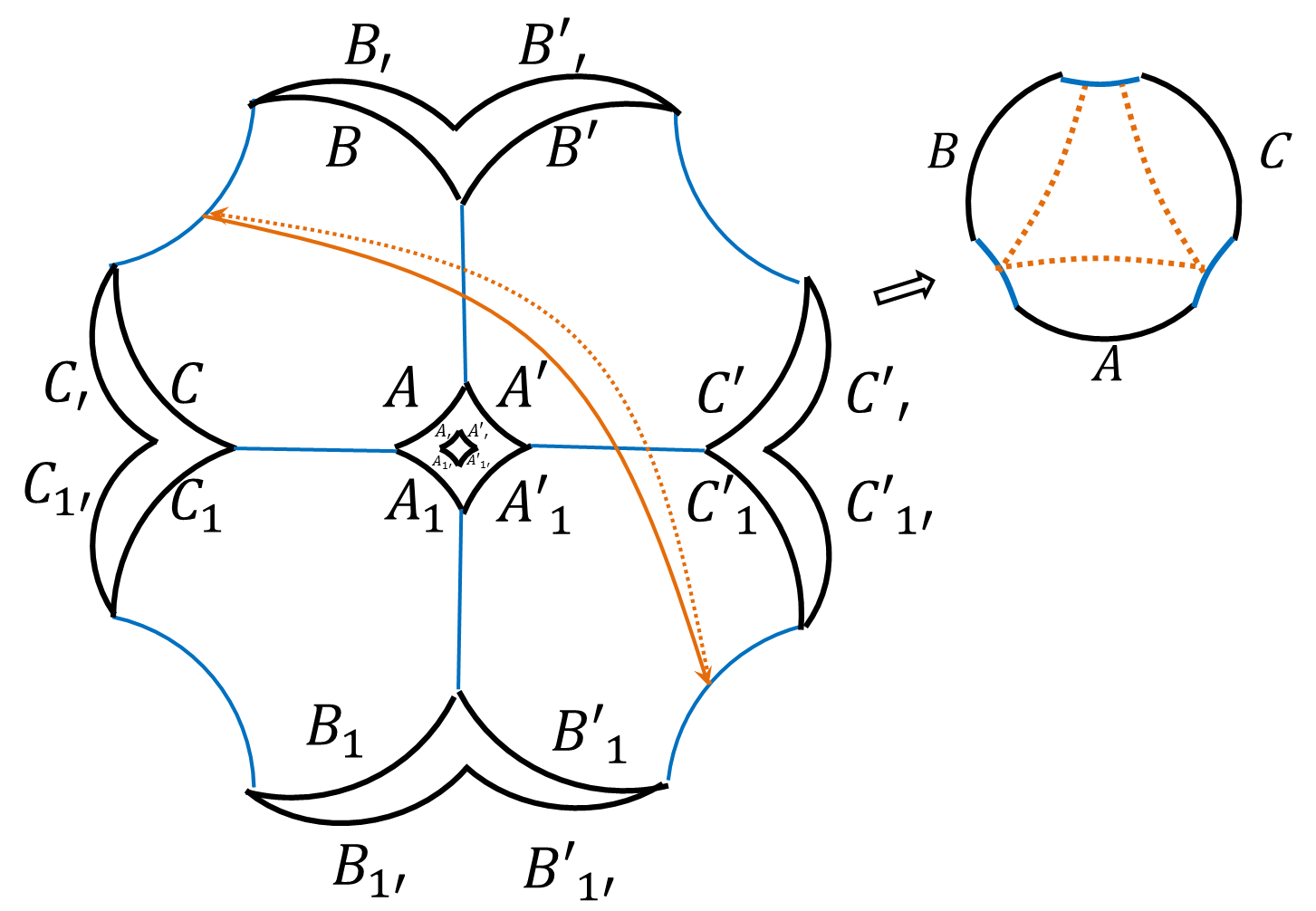}\\
\caption{Canonical purification of $\rho_2$ ($\mathit{left}$) with the minimal cross section denoted by the orange line dual to $S(BB'B_{'}B_{'}'C'C{'}'C_1'C_{1'}')_{\psi_3}$ . This is twice of $\Sigma_{(2)}^{min}(A:B:C)$ denoted by the orange line in the entanglement wedge of $\rho_{ABC}$ ($\mathit{right}$). Notice that $\Sigma_{(2)}^{min}(A:B:C)$ is different from $\Delta_W(A:B:C)$.}
\label{8copy}
\end{figure}

We also draw a pure state in 8-copy canonical purification in Fig.\ref{8copy} where the left orange line denotes a minimal curve which is dual to $S(BB'B_{'}B_{'}'C'C_{'}'C_1'C_{1'}')_{\psi_3}$. It can be seen that this is twice of $\Sigma_{(2)}^{min}(A:B:C)$ defined as the minimal curve with the shape shown in right figure of Fig.\ref{8copy}.

It can be seen that there are some inequalities between cross sections mentioned above, which are
\begin{equation}
E_W(A:BC)+E_W(B:CA)\le \Sigma_{(1)}^{min}(C:A:B)\ ,
\end{equation}
\begin{equation}
\frac{\Sigma_{(1)}^{min}(A:B:C)+\Sigma_{(1)}^{min}(B:C:A)+\Sigma_{(1)}^{min}(C:A:B)}{2}\le \Delta_W(A:B:C)\ ,
\end{equation}
\begin{equation}
\begin{split}
&E_W(A:BC)+E_W(B:CA)+E_W(C:AB)\\
&\le \text{min}\{\Sigma_{(2)}^{min}(A:B:C),\Sigma_{(2)}^{min}(B:C:A),\Sigma_{(2)}^{min}(C:A:B)\}\\
&\le \text{max}\{\Sigma_{(2)}^{min}(A:B:C),\Sigma_{(2)}^{min}(B:C:A),\Sigma_{(2)}^{min}(C:A:B)\}\\
&\le \Delta_W(A:B:C)\ .
\end{split}
\end{equation}

Apart from $2^n$-copy purifications, any even-copy pure state can be constructed (not unique). For example, we can construct 12-copy pure state by tracing out two copies from the 8-copy pure state and then performing canonical purification, i.e., $\psi_4=|\sqrt{\text{Tr}_{A_1B_1C_1A_{'}'B_{'}'C_{'}'}\rho_3}\rangle$.

Regarding that there are many different purifications in this manner, and for each purification there are many different bipartitions (and therefore different entanglement entropies), we deduce that there exist a lot of dual pairs of generalized reflected entropy and its holographic counterpart. We are not going to list all of them and consider them as direct consequence of our discussion above.
\section{Computation of $\Delta_R$ in AdS$_{3}/$CFT$_2$}\label{sec4}

\begin{figure}[htbp]
\centering
\includegraphics[width=4in]{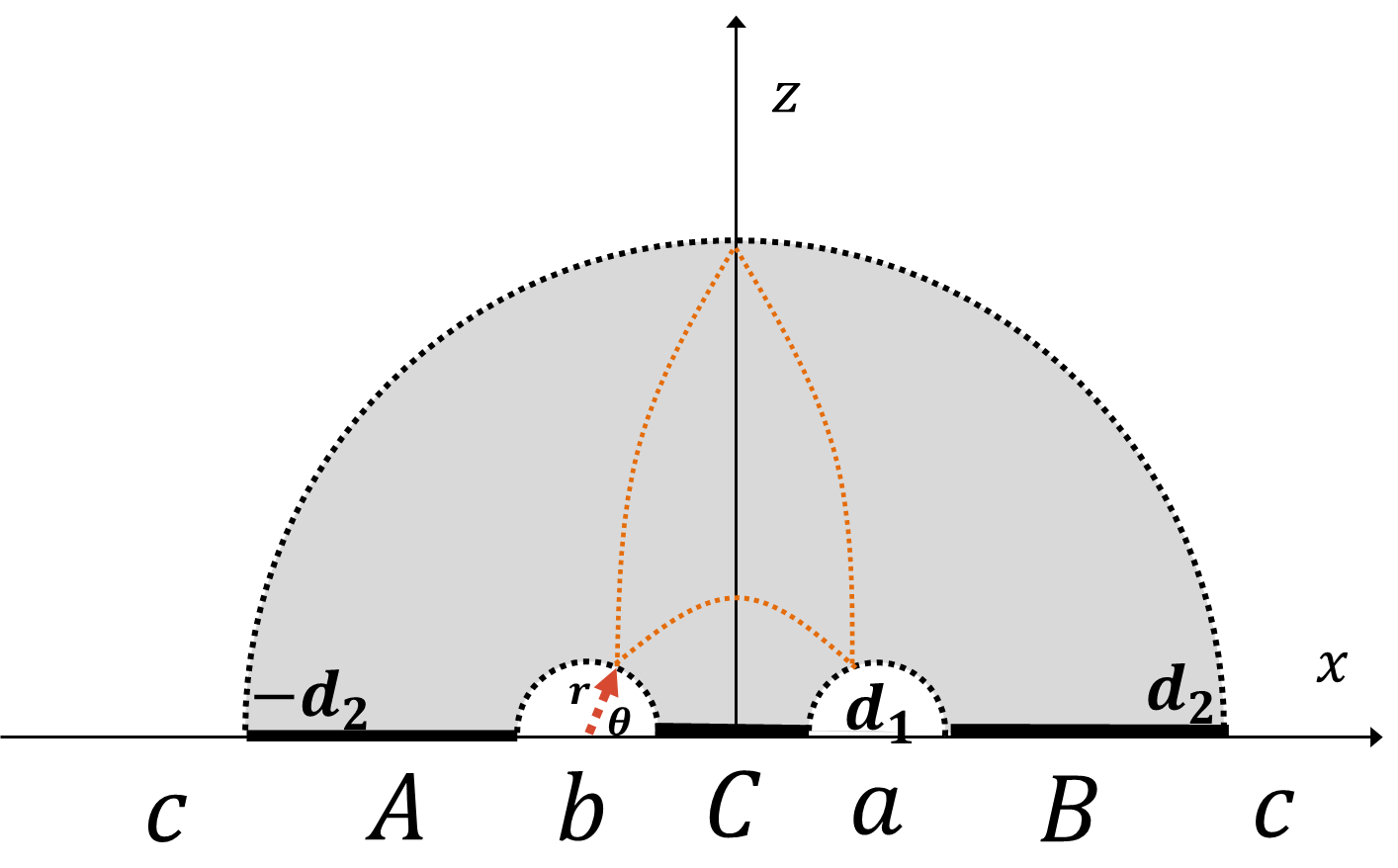}\\
\caption{Three subsystems $A$, $B$ and $C$ in CFT$_2$ and tripartite entanglement wedge cross section in AdS$_3$.}
\label{6pt}
\end{figure}

Now we consider $\Delta_R$ for a simple example in AdS$_{3}$/CFT$_{2}$. We work in Poincar\'e patch, and a static ground state of CFT$_{2}$
on an infinite line is described by a bulk solution with the metric
\begin{equation}
ds^{2}=\frac{dx^{2}+dz^{2}}{z^{2}}\ ,\quad x\in(-\infty,+\infty),z\in[0,+\infty)\ .
\end{equation}
The three subsystems we choose are the intervals $A=[-d_2,-d_1-r]$, $B=[-d_1+r,d_1-r]$,
$C=[d_1+r,d_2]$, where $d_2>d_1>0$ and $r$ is relatively small compared
to both $d_1$ and $d_2$. We require that the entanglement wedge of $ABC$
is connected, as shown in Fig.\ref{6pt}. Let us first consider the holographic computation.
This involves the computation of multipartite entanglement wedge cross section $\Delta_W$ given in~\cite{Umemoto:2018jpc}. In this example we have to find a triangle type configuration with the minimal length, where 3 ending points
of the geodesics are located on 3 Ryu-Takayanagi surfaces (semi-circles) separately, as shown
in Fig.\ref{6pt}. 
Because of the reflection symmetry $x\to-x$, the problem was further reduced to find a special angle $\theta$
such that the length of 3 geodesics is minimal
\begin{equation}
\Delta_{W}(A:B:C)=\min_{\theta}\left[\frac{L(\theta)}{4G_{N}}\right]\ .
\end{equation}

Then we compute $\Delta_R(A:B:C)$ in CFT$_2$ for the same setup in Fig.\ref{6pt} with replica trick.

\begin{figure}[htbp]
           \centering
	\subfigure[]{
	\begin{minipage}[t]{0.45\linewidth}
	\centering
	\includegraphics[width=2.5in]{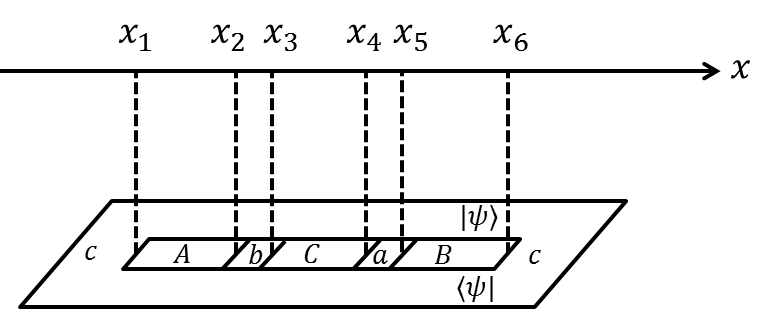}
	\end{minipage}}
	\subfigure[]{
	\begin{minipage}[t]{0.45\linewidth}
	\centering
	\includegraphics[width=2.5in]{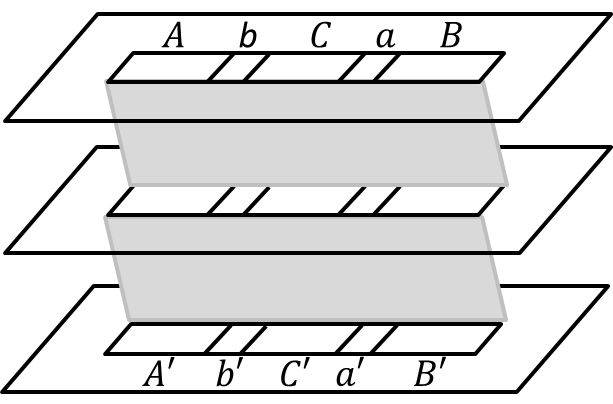}
	\end{minipage}}
	\centering
	\caption{Use replica trick to represent $\psi_1^{(m)}$. (a):  $\text{Tr}_c\rho_0$ and (b): $\psi_1^{(m=6)}$.}
	\label{replica2}
	\end{figure}

We first use replica trick to extend the purification $\psi_3$ to $\psi_3^{(m)}$ following the method in~\cite{Dutta:2019gen}, where $m$ is an even number. The three steps (\ref{stp1}) (\ref{stp2}) (\ref{stp3}) will be generalized to
\begin{equation}
\label{stpi}
\begin{split}
i:\ \psi_1^{(m)}&=|(\text{Tr}_c\rho_0)^{\frac{m}{2}}\rangle\ , \\ 
ii:\ \psi_2^{(m)}&=|(\text{Tr}_{bb'}\rho_1^{(m)})^{\frac{m}{2}}\rangle\ , \\
iii:\ \psi_3^{(m)}&=|(\text{Tr}_{aa'a''a'''}\rho_2^{(m)})^{\frac{m}{2}}\rangle \ .
\end{split}
\end{equation} 
with $\sqrt{\rho}$ changed to $\rho^{\frac{m}{2}}$. These steps can be represented by path integral and replica trick. For instance the first step is illustrated in Fig.\ref{replica2}.

	\begin{figure}[htbp]
	\centering
	\includegraphics[width=6in]{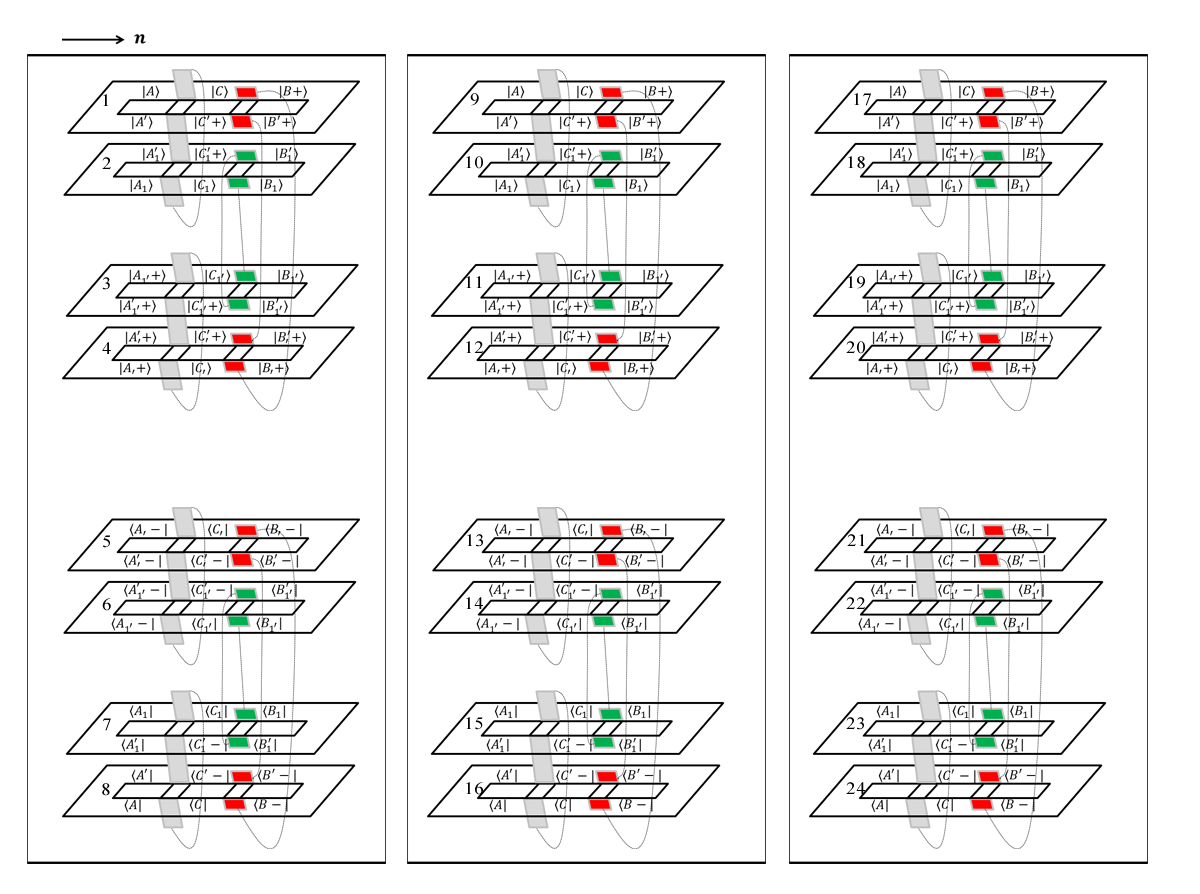}\\
	\caption{Replica trick representing $\text{Tr}_R (\text{Tr}_L\rho_3^{(m)})^{\pmb{n}}$ to calculate $\pmb{n}$-th Renyi entropy. Here $m=2,\pmb{n}=3$ with 24 replicas in total. There are $\pmb{n}$ similar boxes, each representing the density matrix $\rho_3^{(m)}=|\psi_3^{(m)}\rangle \langle \psi_3^{(m)}|$. The rule to glue edges of cut is as follows: in each box, bra and ket of the same object are glued, e.g., the edge $\langle A|$ and $|A\rangle$ are glued. And the ket of $+$ object glues to the bra of the $-$ object in the next box, e.g., $|B+\rangle$ in the left box and $\langle B-|$ in the middle box.}
	\label{replica4}
	\end{figure}

Now we calculate $\pmb{n}$-th R\'enyi entropy
\begin{equation}
\label{ry2}
\begin{split}
S_{\pmb{n}}=\frac{1}{1-\pmb{n}}\ln \frac{\text{Tr}_R(\text{Tr}_L\rho^{(m)}_3)^{\pmb{n}}}{(\text{Tr}\rho^{(m)}_3)^{\pmb{n}}}\ .
\end{split}
\end{equation} 
Compared with (\ref{ry1}), in addition to $\rho_3\to \rho_3^{(m)}$, there is a normalized factor $(\text{Tr}\rho^{(m)}_3)^{\pmb{n}}$.

To compute R\'enyi entropy we have to replicate the previous replicas ($\rho_3^{(m)}$ corresponds to single box in Fig.\ref{replica4} with $m^3$-replica) $\pmb{n}$ times. So there are $m^3\pmb{n}$ replicas in total (shown in Fig.\ref{replica4}), with which we can work out six twist operators $\sigma_i(x_i)$, located at $x_1=-d_2,x_2=-d_1-r,x_3=-d_1+r,x_4=d_1-r,x_5=d_1+r,x_6=d_2$ respectively. It can be counted from replicas that the conformal dimensions $h_i$ of operators $\sigma_i(x_i)$ are (see Appendix \ref{cw})
\begin{equation}
\label{hi}
\begin{split}
h_1=h_6=\frac{c}{24}(m^3-m)\pmb{n} \ , \quad h_2=h_3=\frac{c}{12}(m^2-1)\pmb{n} \ , \quad h_4=h_5=\frac{c}{6}(m-\frac{1}{m})\pmb{n}\ .
\end{split}
\end{equation} 
Although these dimensions look different, they will all go to zero when $m\to 1$. Once twist operators $\sigma_i(x_i)$ are specified, conformal dimensions $h_f$ of the leading operator $\sigma_f$ in OPE contractions $\sigma_i(x_i) \sigma_j(x_j) \to \sigma_f(x_f)$ can also be directly counted. For example,
\begin{equation}
\label{hf}
\begin{split}
h_{16}=h_{23}=h_{45}=\frac{c}{6}(\pmb{n}-\frac{1}{\pmb{n}})\ .
\end{split}
\end{equation} 
It's not surprising that they are equal because $ABC$ are symmetric.

The trace of density matrix is related to 6-point correlation function of twist operators
\begin{equation}
\begin{split}
\text{Tr}_R (\text{Tr}_L\rho_3^{(m)})^{\pmb{n}}=\langle \sigma_1(x_1)\sigma_2(x_2)\sigma_3(x_3)\sigma_4(x_4)\sigma_5(x_5)\sigma_6(x_6)\rangle_{CFT^{\otimes m^3\pmb{n}}}\ .
\end{split}
\end{equation} 
In the large $c$ limit with $\frac{h_i}{c}$ fixed, this correlation function can be determined by a $6$-point Virasoro block $\mathcal{F} $ which in any channel exponentiates~\citep{Hartman:2013mia}
\begin{equation}
\begin{split}
\mathcal{F}  \approx \exp\left[-\frac{c}{6}f\left(\frac{h_f}{c}, \frac{h_i}{c}, x_i\right)\right]
\end{split}
\end{equation} 
where $f$ is determined by the solution of a monodromy problem as follows. Consider the differential equation
\begin{equation}
\label{diff}
\begin{split}
\psi''(z) + T(z) \psi(z) = 0
\end{split}
\end{equation} 
where
\begin{equation}
\begin{split}
T(z) = \sum_{i=1}^6 \left(\frac{6 h_i/c}{(z-x_i)^2}  - \frac{c_i}{z-x_i}\right)\ . 
\end{split}
\end{equation} 
$c_i$ are called accessory parameters restricted by three equations which require $T(z)$ to vanish as $z^{-4}$ at infinity, namely
\begin{equation}
\label{treg}
\begin{split}
\sum_{i=1}^6 c_i  = 0  \ , \quad \sum_{i=1}^6(c_i x_i- \frac{6 h_i}{c})  = 0 \ , \quad \sum_{i=1}^6(c_i x_i^2 - \frac{12 h_i}{c}x_i)  = 0 \ .
\end{split}
\end{equation} 
The differential equation (\ref{diff}) has two solutions, $\psi_1$ and $\psi_2$. As we take the solutions on a closed contour around one or more singular points, they undergo some monodromy
\begin{equation}
\begin{split}
{\psi_1 \choose \psi_2}  \rightarrow M {\psi_1 \choose \psi_2}\ .
\end{split}
\end{equation} 	
	\begin{figure}[htbp]
	\centering
	\includegraphics[width=4in]{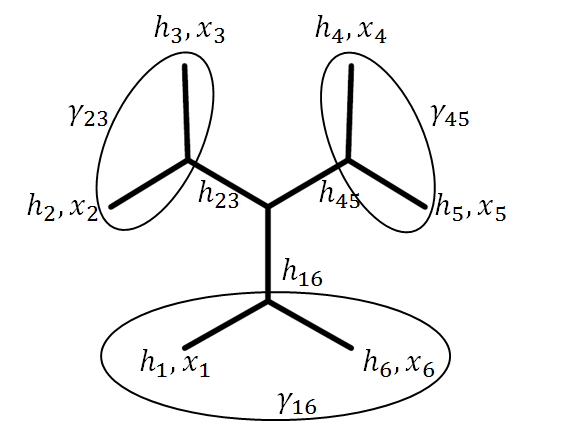}\\
	\caption{The channel and contours chosen to determine the monodromies.}
	\label{cc}
	\end{figure}

We choose contours around the singular points which correspond to the OPE contractions in the chosen channel as shown in Fig.\ref{cc}. That is to say, for each contraction $O_i(x_i) O_j(x_j) \to O_f(x_f)$, we choose a contour $\gamma_f$ enclosing $x_i$ and $x_j$. The monodromies on these cycles should satisfy the conditions
\begin{equation}
\label{trm}
\begin{split}
\text{Tr} M_f = -2 \cos \left(\pi \sqrt{1-\frac{24}{c}h_f}\right)\ .
\end{split}
\end{equation} 
Plus three conditions (\ref{treg}), there are totally six equations of accessory parameters $c_i$. So we can solve $c_i$ which are the partial derivative of $f$ with respect to $x_i$
\begin{equation}
\begin{split}
\frac{\partial f}{\partial x_i} = c_i\ .
\end{split}
\end{equation} 

There are two semiclassical blocks in $\Delta_R(A:B:C)$, namely $f(\frac{1}{6}(\pmb{n}-\frac{1}{\pmb{n}}), h_i, x_i)$ and $f(0, h_i, x_i)$ which are the numerator and the denominator in (\ref{ry2}) respectively. `0' in the later one $f(0, h_i, x_i)$ means that  differential equation (\ref{diff}) has trivial monodromy, i.e., $\text{Tr} M_f =2$. When $m\to 1$, $f(0, 0, x_i)$ becomes constant because it can be easily checked that $c_i=0$ is a solution. Thus, the partial derivatives of $\Delta_R(A:B:C)$ to $x_i$ are
\begin{equation}
\begin{split}
\frac{\partial \Delta_R(A:B:C)}{\partial x_i}=\lim_{m,\pmb{n} \to 1}\frac{1}{1-\pmb{n}}\left[-\frac{c}{3}\frac{\partial f\left(\frac{1}{6}(\pmb{n}-\frac{1}{\pmb{n}}), h_i, x_i\right)}{\partial x_i}\right]\ .
\end{split}
\end{equation} 
Note that only when $r$ is sufficiently less than $d_1$ and $d_1$ is sufficiently less than $d_2$ our channel Fig.\ref{cc} is valid to give the result. Otherwise, $\Delta_R$ will experience phase transitions, as discussed in~\cite{Umemoto:2019jlz}.
         	\begin{figure}[htbp]
           \centering
	\subfigure[]{
	\begin{minipage}[t]{0.95\linewidth}
	\centering
	\includegraphics[width=5in]{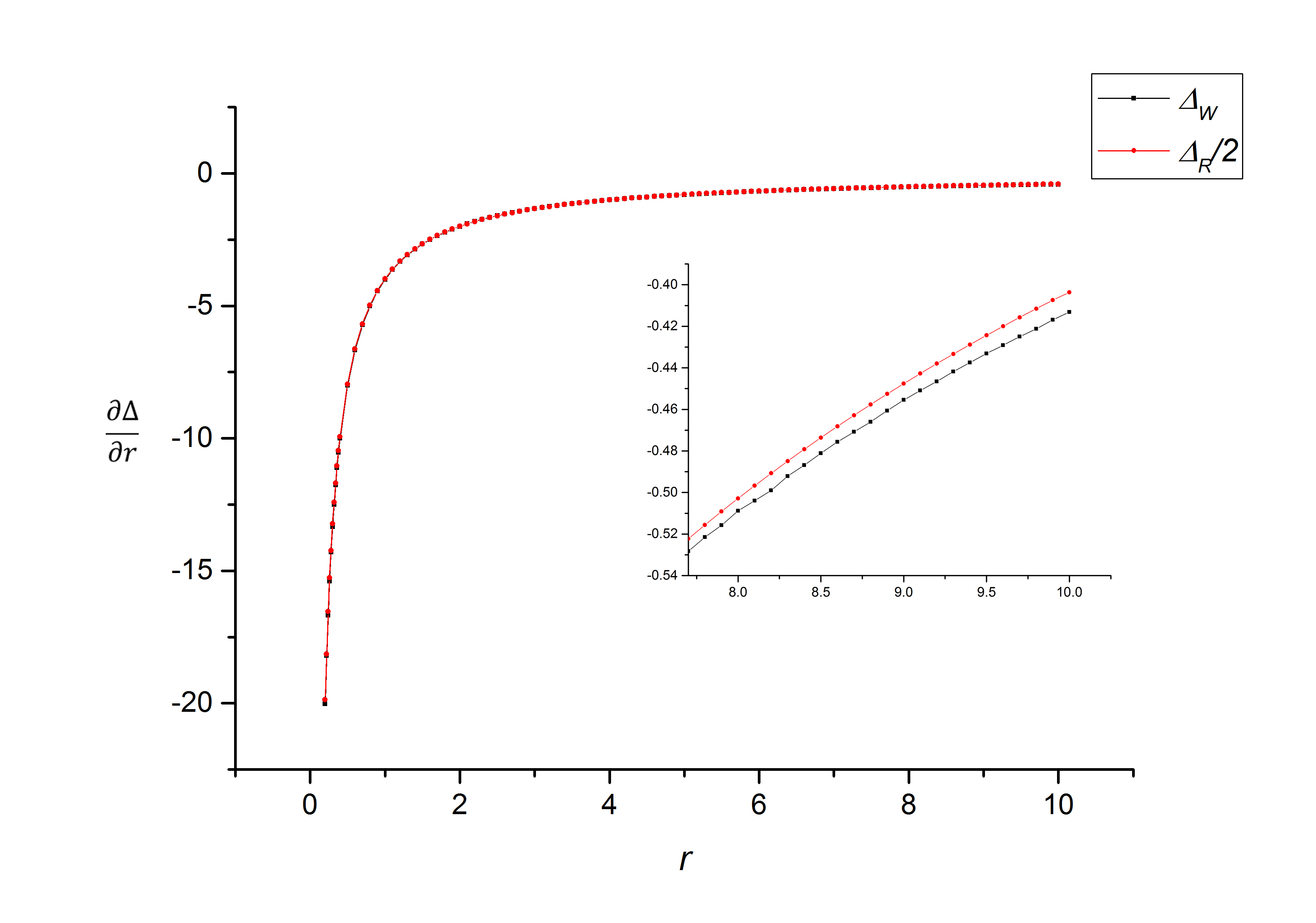}
	\end{minipage}}
	\subfigure[]{
	\begin{minipage}[t]{0.95\linewidth}
	\centering
	\includegraphics[width=5in]{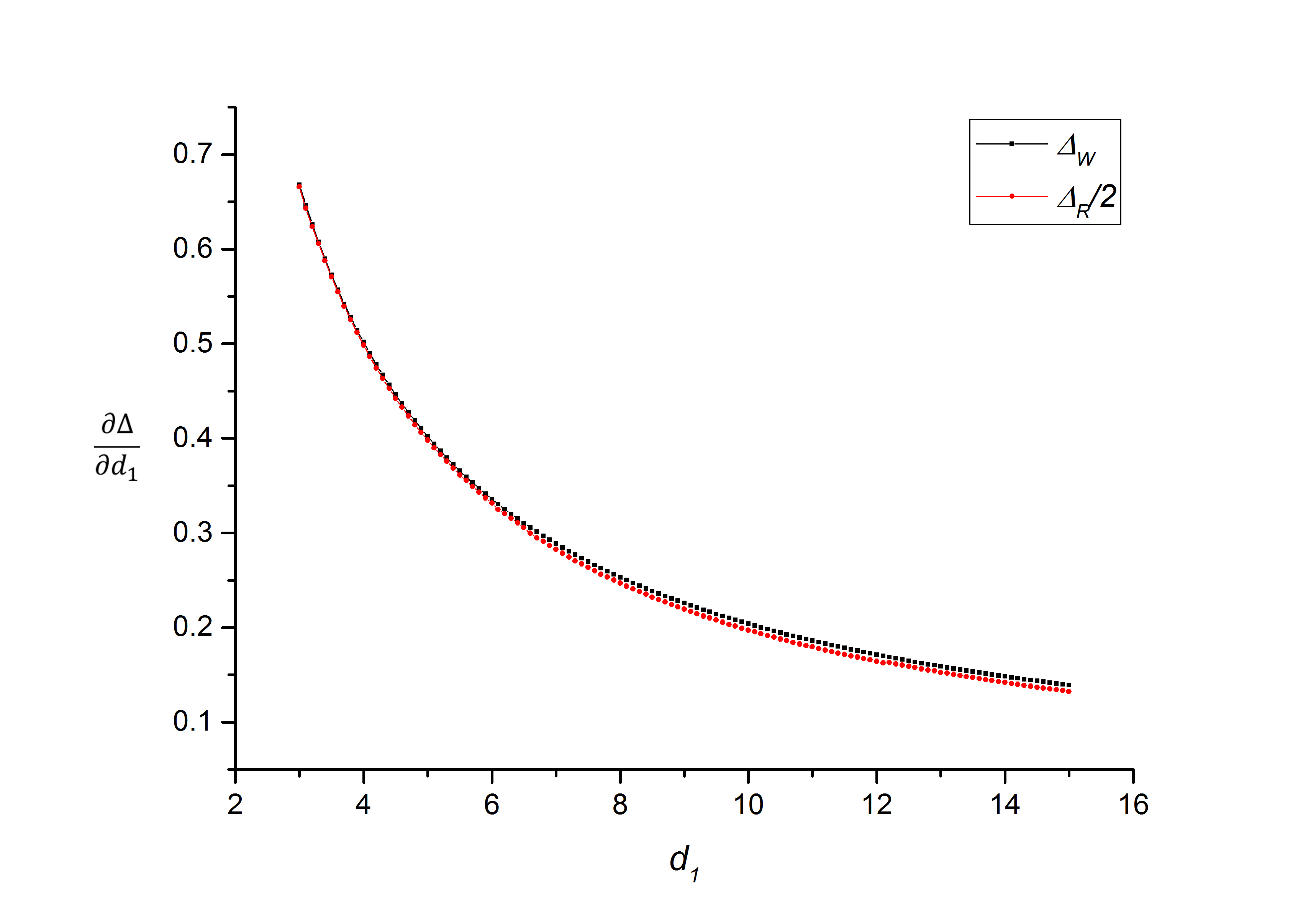}
	\end{minipage}}
	\centering
	\caption{Comparison between the derivative of $\frac{\Delta_R}{2}$ and $\Delta_W$ (divided by $\frac{c}{6}=\frac{1}{4G_N}$) with respect (a): to $r$, with $d_1=20,d_2=100$ and (b): to $d_1$, with $r=0.5,d_2=100$.}
	\label{comp}
	\end{figure}
	
Then the derivative of $\Delta_R$ with respect to $y$ ($y=d_1,d_2 \text{ or }r$) is
\begin{equation}
\begin{split}
\frac{\partial \Delta_R(A:B:C)}{\partial y}&=-\frac{c}{3}\lim_{m,\pmb{n} \to 1}\frac{1}{1-\pmb{n}}\left(\sum_{i=1}^6\frac{\partial f}{\partial x_i}\frac{\partial x_i}{\partial y}\right)\\
&=-\frac{c}{3}\lim_{m,\pmb{n} \to 1}\frac{1}{1-\pmb{n}}\left(\sum_{i=1}^6c_i\frac{\partial x_i}{\partial y}\right)
\end{split}
\end{equation} 
We numerically plot the partial derivative of (half of) $\Delta_R$ with respect to $r$ and $d_1$ and compare it with that of $\Delta_W$ in Fig.\ref{comp}. It can be seen that $\frac{\Delta_R}{2}$ fits well with $\Delta_W$.~\footnote{However, when $r$ or $a$ becomes larger, it can be seen from the numerical data that $\frac{\Delta_R}{2}$ differs gradually from $\Delta_W$.}

\section{Some properties of $\Delta_{R}$}\label{sec5}
In this section we discuss some information theoretic properties of tripartitie reflected entropy $\Delta_{R}$.

When $\rho_{ABC}=|\psi_{ABC}\rangle \langle \psi_{ABC}|$ is pure, $|\psi_1\rangle=|\psi_{ABC}\rangle \otimes |\psi_{A'B'C'}\rangle$, and it can be checked that 
\begin{equation}
\begin{split}
\Delta_{R}(A:B:C)=&EE(AC:B)+EE(A':B'C')+EE(A'_1B_1':C'_1)\\
&+EE(C_{'}:A_{'}B_{'})+EE(B_{1'}C_{1'}:A_{1'})+EE(B'_{1'}:A'_{1'}C'_{1'})\\
=&S(B)+S(A')+S(C'_1)+S(C_{'})+S(A_{1'})+S(B'_{1'})\\
=&2(S(A)+S(B)+S(C))\ .
\end{split}
\end{equation} 

	\begin{figure}[htbp]
	\centering
	\includegraphics[width=3in]{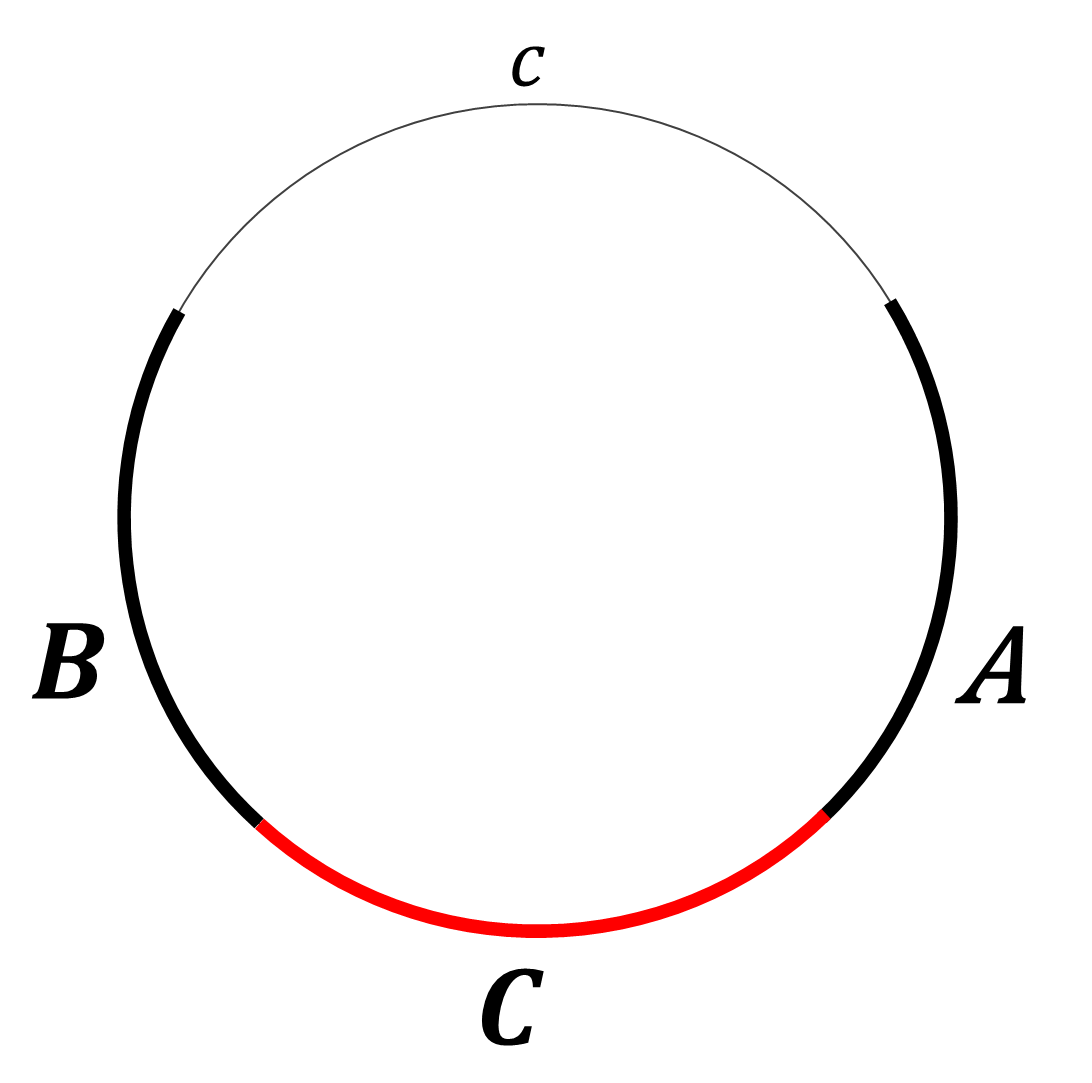}\\
	\caption{A special case in which we study quantum information aspects of $\Delta_{R}(A:B:C)$.}
	\label{sc}
	\end{figure}
	
For some special $\rho_{ABC}$ as shown in Fig.\ref{sc} where $a,b\to 0$, one can easily check some properties of $\Delta_R(A:B:C)$. The purification is
\begin{equation}
\begin{split}
|\psi_3\rangle=|\psi_1\rangle_{ABCA'B'C'}\otimes |\psi_1\rangle_{A_1B_1C_1A_1'B_1'C_1'}\otimes |\psi_1\rangle_{A_{'}B_{'}C_{'}A'_{'}B'_{'}C'_{'}}\otimes |\psi_1\rangle_{A_{1'}B_{1'}C_{1'}A_{1'}'B_{1'}'C_{1'}'}
\end{split}
\end{equation} 
where $|\psi_1\rangle=|\rho_{ABC}^{\frac{1}{2}}\rangle$. Then
\begin{equation}
\label{qidr}
\begin{split}
\Delta_{R}(A:B:C)\equiv &EE(AA'A_1A'_1B_1B'_1B_{1'}B'_{1'}CC_{'}C_1C_{1'}:A_{'}A'_{'}A_{1'}A'_{1'}BB'B_{'}B'_{'}C'C'_{'}C'_1C'_{1'})\\
=&EE(AA'C:BB'C')+EE(A_1A'_1B_1B_1'C_1:C'_1)\\
&+EE(C_{'}:A_{'}A'_{'}B_{'}B'_{'}C'_{'})+EE(B_{1'}B'_{1'}C_{1'}:A_{1'}A'_{1'}C'_{1'})\\
=&S(AA'C)+S(C'_1)+S(C_{'})+S(A_{1'}A'_{1'}C'_{1'})\\
=&2(S(AA'C)+S(C))
\end{split}
\end{equation} 
where $S(X)$ means $S(\text{Tr}_{X}\rho_1)$. From (\ref{qidr}) we can see that $\Delta_{R}(A:B:C)\ge S(C)>0$, which implies that $\rho_{ABC}$ in this case can't be seperable.

Now we check some properties of $\Delta_{R}(A:B:C)$ in this case. Due to the positivity of mutual information
\begin{equation}
\begin{split}
I(AC:A')=S(AC)+S(A')-S(AA'C)\ge 0
\end{split}
\end{equation} 
it can be derived that
\begin{equation}
\begin{split}
\Delta_{R}(A:B:C)=&2(S(AA'C)+S(C))\\
\le&2(S(AC)+S(A')+S(C))\\
=&2(S(AC)+S(A)+S(C))\ .
\end{split}
\end{equation} 
Similarly
\begin{equation}
\label{ub}
\begin{split}
\Delta_{R}(A:B:C)\le2(S(BC)+S(B)+S(C))\ .
\end{split}
\end{equation} 
We could not show
\begin{equation}
\begin{split}
\Delta_{R}(A:B:C)\le2(S(AB)+S(A)+S(B))\ ,
\end{split}
\end{equation} 
but since the right side has much larger UV divergence, this is expected to be true.~\footnote{We thank Koji Umemoto for pointing this out.}

From strong sub-additivity
\begin{equation}
\begin{split}
S(BB'C')+S(ABC)\ge S(B'C')+S(AC)
\end{split}
\end{equation} 
and Araki-Lieb inequality
\begin{equation}
\begin{split}
S(AB)-S(C)\le S(ABC)\ ,
\end{split}
\end{equation} 
it can be derived that 
\begin{equation}
\label{I}
\begin{split}
\Delta_{R}(A:B:C)=&2(S(BB'C')+S(C))\\
\ge&2(S(C)+ S(B'C')+S(AC)-S(ABC))\\
\ge&2(S(AB)+S(BC)+S(AC)-2S(ABC))\ .
\end{split}
\end{equation} 
which is defined as $D_3(A:B:C)\times 2$ in~\cite{Umemoto:2019jlz}.

We can also derive polygamy, namely for a pure state $\rho_{A_1A_2BC}$
\begin{equation}
\begin{split}
\Delta_{R}(A_1A_2:B:C)=&2(S(A_1A_2)+S(B)+S(C))\\
=&2(S(BC)+S(B)+S(C))\\
\le&2(S(B)+S(C)+S(B)+S(C))\\
=&2(S(B)+S(C)+S(B)+S(C)+S(A_1)-S(A_2BC)+S(A_2)-S(A_1BC))\\
=&2(S(B)+S(C)+S(A_1)-S(A_1BC)+S(B)+S(C)+S(A_2)-S(A_2BC))\\
\le&\Delta_{R}(A_1:B:C)+\Delta_{R}(A_2:B:C)
\end{split}
\end{equation} 
where in the third line we used the positivity of mutual information $I(B:C)$ and in the last line we used (\ref{I}).

From strong sub-additivity and positivity of mutual information
\begin{equation}
\begin{split}
S(AA'C)+S(BB'C)\ge& S(AA')+S(BB')\\
I(C:C')\ge&0\ ,
\end{split}
\end{equation} 
it can be derived that
\begin{equation}
\begin{split}
\Delta_{R}(A:B:C)=&2(S(AA'C)+S(C))\\
=&S(AA'C)+S(BB'C')+2S(C)\\
=&S(AA'C)+S(BB'C)+S(C)+S(C')\\
\ge &S(AA')+S(BB')+S(CC')\\
=&S_R(A:BC)+S_R(B:CA)+S_R(C:AB)\ .
\end{split}
\end{equation} 

\section{Conclusion}\label{sec6}

In this paper, we defined a class of generalized reflected entropy for multipartite states. We show that the generalizations of reflected entropy can be defined canonically. We particularly show that the generalization of reflected entropy to multipartite case is not unique. After $n$ steps of canonical purifications we have obtained a big pure state associated to $2^n$ copies of the original Hilbert space. Each bipartition of the large Hilbert space will define a generalized reflected entropy. In this sense, the generalization depends on both $n$ and the bipartition. Based on this one can construct pure state in any even copies of Hilbert spaces. We develop a general method using replica trick and twist operators in CFTs to compute generalized reflected entropies.
 
Based on the holographic conjecture of reflected entropy \citep{Dutta:2019gen}, we defined a class of minimal surfaces $\Sigma^{min}$ as the holographic counterparts of the generalized reflected entropies, and in particular we show that for holographic theories there is a one to one correspondence between generalized reflected entropy and $\Sigma^{min}$. It leads us to propose a new class of entropies in CFT as dual of various combinations of cross-sections in the entanglement wedge and therefore discovered a new class of quantities which can be used to test AdS/CFT. In tripartite case we focus on a particular generalized reflected entropy $\Delta_{R}(A:B:C)$ and show that its holographic dual is twice of the multipartite entanglement wedge cross sections $\Delta_W$.
We explicitly computed $\Delta_{R}$ for a simple setup in AdS$_{3}$/CFT$_{2}$ and find precise agreement with holographic computation.

Several future questions are in order: First, generalize our holographic conjectures to black hole backgrounds and time-dependent background geometry. Second, generalize our new entropy measures systematically to $n$-partite case and to higher dimensions where we expect that many new types of generalized reflected entropy will appear following our construction. Third,
looking for the dictionary between generalized reflected entropy and minimal cross-sections in $n$-partite case. Understanding holographic $n$-partite states will be quite useful to understand the emergence of bulk geometry from boundary CFT.
We shall report the progress in future publications.

\section*{Acknowledgements}

We thank Koji Umemoto for a lot of helpful discussions.

\appendix
\section{Derivation of conformal weights}\label{cw}

Let's first derive the conformal dimensions in $m=2, \pmb{n}=3$ case as an appetizer. Replica trick shown in Fig.\ref{replica4} is determined by six twist operators $\sigma_i (x_i)$ which are made up of 5 nontrivial operators connecting respectively 5 intervals of replicas, namely $\Sigma_A, \Sigma_b,\Sigma_C,\Sigma_a,\Sigma_B$ from left to right. We mark the 24 replicas by integers from 1 to 24. These operators can be described by representation of cyclic groups. For example, $(123)$ means that the lower edge of cut in replica 1 is glued to the upper edge of cut in replica 2, the lower edge of cut in replica 2 is glued to the upper edge of cut in replica 3, and the lower edge of cut in replica 3 is glued to the upper edge of cut in replica 1. In this way, we can read from Fig.\ref{replica4} that
\begin{equation}
\begin{split}
\Sigma_A&=(1,8)(2,7)(3,14)(4,13)(9,16)(10,15)(11,22)(12,21)(17,24)(18,23)(19,6)(20,5)\\
\Sigma_b&=(1,2)(3,4)(5,6)(7,8)(9,10)(11,12)(13,14)(15,16)(17,18)(19,20)(21,22)(23,24)\\
\Sigma_C&=(1,16,9,24,17,8)(2,7,18,23,10,15)(3,14,11,22,19,6)(4,5,20,21,12,13)\\
\Sigma_a&=(1,4)(2,3)(5,8)(6,7)(9,12)(10,11)(13,16)(14,15)(17,20)(18,19)(21,24)(22,23)\\
\Sigma_B&=(1,16)(2,7)(3,6)(4,13)(9,24)(10,15)(11,14)(12,21)(17,8)(18,23)(19,22)(20,5)\ .
\end{split}
\end{equation} 
Then,
\begin{equation}
\begin{split}
\sigma_1&=\Sigma_A=(1,8)(2,7)(3,14)(4,13)(9,16)(10,15)(11,22)(12,21)(17,24)(18,23)(19,6)(20,5)\\
\sigma_2&=\Sigma_A^{-1} \Sigma_b=(1,7)(2,8)(3,13)(4,14)(5,19)(6,20)(9,15)(10,16)(11,21)(12,22)(17,23)(18,24)\\
\sigma_3& =\Sigma_b^{-1}\Sigma_C=(8,2)(1,15)(16,10)(9,23)(24,18)(17,7)(3,13)(14,12)(11,21)(19,5)(6,4)(22,20)\\
\sigma_4 &=\Sigma_C^{-1} \Sigma_a=(1,13)(4,8)(2,6)(3,15)(5,17)(7,19)(9,21)(12,16)(10,14)(11,23)(20,24)(18,22)\\
\sigma_5 &=\Sigma_a^{-1} \Sigma_B=(2,6)(7,3)(1,13)(16,4)(10,14)(15,11)(9,21)(24,12)(18,22)(23,19)(17,5)(8,20)\\
\sigma_6 &=\Sigma_B^{-1}=(1,16)(2,7)(3,6)(4,13)(9,24)(10,15)(11,14)(12,21)(17,8)(18,23)(19,22)(20,5)
\end{split}
\end{equation} 
so that conformal weights are $h_i=\frac{c}{24}(2-\frac{1}{2})12$, which can be checked by (\ref{hi}). And the leading operators $\sigma_f$ in OPE contractions $\sigma_i(x_i) \sigma_j(x_j) \to \sigma_f(x_f)$ are
\begin{equation}
\begin{split}
\sigma_{16}&=\sigma_1\sigma_6=\Sigma_A\Sigma_B^{-1}=(1,9,17)(3,19,11)(6,14,22)(8,24,16)\\
\sigma_{23}&=\sigma_2\sigma_3=\Sigma_A^{-1}\Sigma_C=(1,9,17)(4,20,12)(6,14,22)(7,23,15)\\
\sigma_{45}&=\sigma_4\sigma_5=\Sigma_C^{-1}\Sigma_B=(3,19,22)(4,12,20)(7,15,23)(8,24,16)
\end{split}
\end{equation} 
so that conformal weights are $h_f=\frac{c}{24}(3-\frac{1}{3})4$, which can be checked by (\ref{hf}).

Now we are ready to generalize our derivation to $m,\pmb{n}>2$ case. From above we can see that twist operator $\sigma_i$ contain $m$-cycles and $\sigma_f$ contain $\pmb{n}$-cycles, i.e.,
\begin{equation}
\begin{split}
h_i&\propto m-\frac{1}{m}\\
h_f&\propto \pmb{n}-\frac{1}{\pmb{n}}\ .
\end{split}
\end{equation} 
When $m,\pmb{n}\to \infty$, 
\begin{equation}
\begin{split}
h_1,h_6&\to \frac{c}{24}m^3\pmb{n}\\
h_2,h_3&\to \frac{c}{24}2m^2\pmb{n}\\
h_4,h_5&\to \frac{c}{24}4m\pmb{n}\\
h_f&\to  \frac{c}{24}4\pmb{n}
\end{split}
\end{equation} 
because it can be checked that cyclic groups of twist operators $\sigma_1,\sigma_2,\sigma_3,\sigma_4,\sigma_5,\sigma_6,\sigma_{16},\sigma_{23},\sigma_{45}$ pass through $m^3\pmb{n},2m^2\pmb{n},2m^2\pmb{n},4m\pmb{n},4m\pmb{n},m^3\pmb{n},4\pmb{n},4\pmb{n},4\pmb{n}$ replicas respectively. Therefore,
\begin{equation}
\begin{split}
h_1,h_6&= \frac{c}{24}(m-\frac{1}{m})m^2\pmb{n}\\
h_2,h_3&= \frac{c}{24}(m-\frac{1}{m})2m\pmb{n}\\
h_4,h_5&= \frac{c}{24}(m-\frac{1}{m})4\pmb{n}\\
h_f&=  \frac{c}{24}(\pmb{n}-\frac{1}{\pmb{n}})4 \ .
\end{split}
\end{equation}

\end{document}